\documentclass{emulateapj}
\usepackage{apjfonts}
\usepackage{graphicx}
\usepackage{amsmath}
\usepackage{natbib}
\usepackage{epsfig}
\newcommand{\etal}{{\it et al.~}}

\def\la{\mathrel{\mathchoice 
{\vcenter{\offinterlineskip\halign{\hfil$\displaystyle##$\hfil\cr<\cr\sim\cr}}}
{\vcenter{\offinterlineskip\halign{\hfil$\textstyle##$\hfil\cr<\cr\sim\cr}}}
{\vcenter{\offinterlineskip\halign{\hfil$\scriptstyle##$\hfil\cr<\cr\sim\cr}}}
{\vcenter{\offinterlineskip\halign{\hfil$\scriptscriptstyle##$\hfil\cr<\cr\sim\cr}}}}}
\def\ga{\mathrel{\mathchoice 
{\vcenter{\offinterlineskip\halign{\hfil$\displaystyle##$\hfil\cr>\cr\sim\cr}}}
{\vcenter{\offinterlineskip\halign{\hfil$\textstyle##$\hfil\cr>\cr\sim\cr}}}
{\vcenter{\offinterlineskip\halign{\hfil$\scriptstyle##$\hfil\cr>\cr\sim\cr}}}
{\vcenter{\offinterlineskip\halign{\hfil$\scriptscriptstyle##$\hfil\cr>\cr\sim\cr}}}}}

\begin{document}

 \title{Reconstructing the equation of state and  density parameter for dark energy from combined analysis of recent SNe Ia, OHD and BAO data}

 \author{Debabrata Adak}
\affil{Astroparticle Physics and Cosmology Division, Saha Institute of Nuclear Physics, 1/AF Bidhannagar, Kolkata 700064, India}
\email{debabrata.adak@saha.ac.in}

 \author{Abhijit Bandyopadhyay}
\affil{RKM Vivekananda University, Belur Math, Howrah 711202, India}
\email{abhi.vu@gmail.com}

 \author{Debasish Majumdar}
\affil{Astroparticle Physics and Cosmology Division, Saha Institute of Nuclear Physics, 1/AF Bidhannagar, Kolkata 700064, India}
\email{debasish.majumdar@saha.ac.in}
 
\thispagestyle{empty}

\sloppy

\begin{abstract}
We adopt a model independent method to reconstruct the dark energy equation of state by analyzing 5 sets of SNe Ia data along with Baryon Acoustic Oscillation (BAO) and Observational Hubble Data (OHD). The SNe Ia data sets include the most recent UNION2  data and other data compilations from the year 2007 to the present. We assume a closed form parametrization of the luminosity distance in terms of redshift and perform a  $\chi^2$ analysis of the observational data. The matter density at the present epoch $\Omega_m^0$ is also taken to be a parameter in the analysis and its best-fit values are obtained for each of the data sets. We found a strong dependence of dark energy equation of state on the matter density in the present and earlier epoch. From the analysis, we also predict the lower limit of matter density parameter at an earlier epoch within 1$\sigma$ confidence level for a flat FRW universe. The dark energy equation of state appears to be a slow varying function of $z$. The variation of dark energy density parameter and the matter density parameter are also shown along with their 1$\sigma$ variations.
\end{abstract}
\vskip 1cm


\maketitle

\section{Introduction}
\label{sec:intro}
Observations on type Ia supernovae (SNe Ia) \citep{sn1, sn2} during nearly last two  decades reveal that the universe is undergoing accelerated expansion in the present epoch. This accelerated expansion can be explained by invoking the existence of dark energy - a hypothetical energy component with a negative pressure. Despite several past and ongoing efforts the nature and origin of `dark energy' remains a mystery. The evidence of dark energy can also be predicted from the observation of Baryon Acoustic Oscillation (BAO) \citep{bao}, Hubble data based on differential ages of the galaxies (OHD) \citep{ohd,ma} etc. 

Some of the properties of dark energy can be extracted by performing analysis of the observational data. In general, there are different approaches for the analysis of SNe Ia data (obtained in the form of luminosity distance modulus versus redshift). One of the approaches involves choice of some arbitrary parametrization of the dark energy equation of state $w_X(z)$ (=$p/\rho$). $p$ and $\rho$ respectively denoting the pressure and energy density associated with the dark energy. The luminosity distance $d_L(z)$ and the bolometric magnitude $\mu(z)$ at redshift $z$ are then found by using the assumed parametric form of $w_X(z)$. Such approaches are taken up and discussed in detail in \citep{mth1,mth2,mth3,mth4,mth5,mth6,mth7,mth8,mth9,mth10,mth11,mth12,mth13,mth14,
mth15,mth16,mth17,mth18,mth19,mth20}. 
In another kind of approach $\mu(z)$ are first fitted with observational data and then one finds the dark energy equation of state $w_X(z)$ \citep{deb3,deb7,deb5,deb2,deb8,deb1}.

In this paper we have considered different compilations of SNe Ia data sets \textit{viz.} 
\citep{Riess07,WoodVasey,Davis,kowalski,Kessler} and \citep{Amanullah}. From SNe Ia observations these groups have tabulated the values of $\mu(z)$ for different values of redshift ($z$) within the limit 0.001 $\leq$ $z$ $\leq$ 1.76. We have considered a parametric form of $d_L(z)$ and expressed $\mu(z)$ in terms of this parametric form of $d_L(z)$. We make a $\chi^2$ analysis of the combined data sets of SNe Ia, BAO and OHD to obtain the best-fit values of the parameters from the observational data. We have taken 5 different SNe Ia data sets (described later) and for the combined analysis each of these data sets are combined with BAO and OHD data. In this work the matter density at the present epoch $(\Omega_m^0)$ is also taken to be a parameter for the $\chi^2$ analysis, and by performing the $\chi^2$ minimization we obtain the best-fit value of  $(\Omega_m^0)$ along with the parameters appearing in the parametrization of $d_L(z)$. Having thus obtained $d_L(z)$ as a function of $z$ we calculate the variation of $\omega_X(z)$ as a function of $z$ for the best-fit values of the parameters and their $1 \sigma$ limits as well for each of the 5 data sets considered. In all the calculations we assume that present universe is spatially flat and contains only matter and dark energy. The results of the analysis show that knowledge of the matter density of the universe at some earlier epoch is instrumental in providing observational evidences in favour of varying dark energy or cosmological constant solutions. We have also shown the simultaneous variation of matter density parameter $\Omega_m(z)$ and dark energy density parameter $\Omega_X(z)$ with $z$ for the best-fit values of the parameters (obtained from $\chi^2$ fitting)  and their $1 \sigma$ range. We also found the epoch at which the dark energy  started dominating over the matter component of the universe.

The paper is organized as follows. In Section\ \ref{sec:reconstruction} the formalism for reconstruction of dark energy is described. The methodology of the analysis are given in Section\ \ref{sec:analysis}. In Section\ \ref{sec:results} we describe results of  analysis of different sets of data and explain our results. Finally in Section\ \ref{sec:conclusion} we make some concluding remarks.

\section{Reconstruction of the equation of state of dark energy}
\label{sec:reconstruction}
In standard FRW cosmology, for a spatially flat universe, the luminosity distance $d_L(z)$ of an object at a redshift $z$ is related to the  Hubble parameter
$H(z)$ as
\begin{eqnarray}
H(z)
\equiv  \frac{\dot{a}}{a} 
= c\left[\frac{d}{dz}\left(\frac{d_L(z)}{1+z}\right)\right]^{-1}
\label{hub1}
\end{eqnarray}
where $c$ is the velocity of light and $a$ is the scale factor, whose time evolution determines the time evolution of the universe. Modeling the total content (other than gravitational field) of the universe as a perfect fluid characterized by its energy density $\rho$ and pressure $p$, the (00) component of the Einstein's equation for a spatially flat universe ($K=0$) gives
\begin{eqnarray}
H^2(z)&=& \frac{8\pi G}{3}\rho(z) \label{hub2}\,\,\,\,.
\end{eqnarray}
Observations from WMAP experiment suggest that the present universe is spatially flat and contribution of radiation to the total density of the universe in negligible \citep{komatsu}. Measured values of redshifts of Supernova Ia events ($z < 2$) correspond to epochs close to the present epoch in cosmological time scale. Therefore for analysis of Supernova Ia observations  we are permitted to use  Eqs.\ (\ref{hub1}) and (\ref{hub2}). In this context we can also write the total energy density of the universe by neglecting the radiation energy density as $\rho(z) = \rho_m(z) + \rho_X(z)$ where $\rho_m(z)$ and $\rho_X(z)$ denote the contribution in the energy density due to (non-relativistic) matter and dark energy. This can also be expressed in terms of the corresponding density parameters $\Omega_m  \equiv (3H^2/8\pi G)\rho_m$ and $\Omega_X \equiv (3H^2/8\pi G)\rho_X$ as $\Omega_m(z) + \Omega_X(z) = 1$. Assuming an effective equation of state for the dark energy $w_X(z) = \rho_X(z)/p_X(z)$, the Hubble parameter can be expressed as
\begin{eqnarray}
\frac{H^2(z)}{H_0^2} 
&=& \Omega_m^0(1+z)^3+\Omega_X^0\exp\left(\int_0^{z} 
3(1+w_X(z^\prime))\frac{dz^\prime}{1+z^\prime}\right)\, ,
\label{hub3}
\end{eqnarray}
where $H_0$ is the value of the Hubble parameter at the present epoch. $\Omega_m^0$ and $\Omega_X^0$ respectively denote values of matter and dark energy density parameter at the present epoch. Differentiating both sides of equation Eq.\ (\ref{hub3}) with respect to $z$ we obtain the equation of state of dark energy  as
\begin{eqnarray}
w_X (z) 
&=& -1 + \displaystyle\left[\frac{\frac{2}{3}\frac{(1+z)}{H(z)}\frac{dH(z)}{dz}-
\Omega_m^0\frac{(1+z)^3}{H^2(z)/H_0^2}}{1-\Omega_m^0\frac{(1+z)^3}
{H^2(z)/H_0^2}}\right]\,\,.
\label{eos}
\end{eqnarray}
We use the above equation to reconstruct the dark energy equation of state from observational data. Such a reconstruction therefore requires extraction of the quantities $H(z)$ and $\Omega_m^0$ from the observational data. From the measured values of the luminosity distances $(d_L(z))$ of type Ia Supernovae at different redshifts $(z)$ we can obtain  $H(z)$ at different redshifts using Eq.\ (\ref{hub1}). The observational Hubble Data (OHD) based on the differential ages of the galaxies also provide values of the Hubble parameter $H(z)$ at some redshift $z$ values. On the other hand, the measurement of Baryon Acoustic Oscillations (BAO) from the study of large-scale correlation function of sky surveys of several thousands of luminous red galaxies provides a value for the quantity \citep{bao,zhang}
\begin{eqnarray}
A(z_1) &=& \frac{\sqrt{\Omega_m^0}}{{[H(z_1)/H_0]}^{1/3}} \left[\frac{1}{z_1} \int_0^{z_1} \frac{dz}{H(z)/H_0}\right] ^{2/3}\label{constructedA}
\end{eqnarray}
with $z_1 = 0.35$.  Eq.\ (\ref{constructedA}) thus connects the $H(z)$ and $\Omega_m^0$ to the observed quantity $A(z=0.35)$. A joint analysis of the SNe Ia, OHD and BAO data can therefore simultaneously constrain the parameters $H(z)$ and $\Omega_m^0$.

The Hubble parameter $H(z)$ and its variation with $z$ show up in different ways in the SNe Ia, OHD and BAO observations. The Hubble parameter is again related to the luminosity distance, $d_L(z)$ by Eq.\ (\ref{hub1}). Choice of some parametric form of $d_L(z)$ therefore relates the observed quantities with the parameters of $d_L(z)$. In this work we use the parametrization of the luminosity distance as
\begin{eqnarray}
d_L(a,b;z) &=& \frac{c}{H_0} \left[\frac{z(1+az)}{1+bz}\right]
\label{dlzparam}
\end{eqnarray}
which respect the conditions that $d_L=0$ at $z=0$ and $d_L \propto z$ for large $z$ corresponding  to the radiation dominated era. SNe Ia data are available for $z \la 1.76$ - an era which is dominated
by matter and dark energy. Using this parametric form and Eq.\ (\ref{hub1})  we can express the Hubble parameter and its $z$-derivative  in terms of parameters $a$ and $b$ as
\begin{eqnarray}
H(a,b;z) 
&=& 
H_0 \left[\frac{(1+z)^2 (1+bz)^2}{1 + 2az + (a - b + ab)z^2}\right]
\label{habz}
\end{eqnarray}
The expression for $H(a,b;z)$  thus obtained can be used to express the quantity $A$ (in Eq.\ (\ref{constructedA})) in terms of $a$ and $b$ as
\begin{eqnarray}
A(a,b,\Omega_m^0) &=& \frac{\sqrt{\Omega_m^0}}{{[H(a,b;z_1)/H_0]}^{1/3}} \left[\frac{1}{z_1} \int_0^{z_1} \frac{dz}{H(a,b;z)/H_0}\right] ^{2/3}
\label{Aabomega}
\end{eqnarray}
with $z_1=0.35$,
where $\Omega_m^0$, along with $a$ and $b$, will be treated as parameters to be determined from observation. The method of analysis of observational data for obtaining the parameters $a$, $b$ and $\Omega_0$ is discussed in Sec.\ \ref{sec:analysis}. Finally the dark energy equation of state $w(z)$ as given in Eq.\ (\ref{eos}) can be expressed in terms of the parameters $a$, $b$ and $\Omega_m^0$ as
\begin{eqnarray}
w_X(a,b,\Omega_m^0; z) = 
\frac{\frac{4}{3(1+bz)}
\left[1+b+2bz-\frac{(1+z)(1+bz)(a+(ab+a-b)z)}{1+2az+(ab+a-b)z^2}\right]-1}
{1-\Omega_m^0\frac{[1+2az+(ab+a-b)z^2]^2}{(1+z)(1+bz)^4}}\nonumber\\
\label{wxab}
\end{eqnarray}
The best-fit values of the parameters $a$, $b$ and $\Omega_m^0$ obtained from the analysis of the experimental data can be used in Eq.\ (\ref{wxab}) to obtain the variation of the equation state of dark energy with redshift.
 
Also we note that Eq.\ (\ref{hub3}) can be identified with the equation $\Omega_m(z) + \Omega_X(z) = 1$ with 
\begin{eqnarray}
\Omega_m(z) &=& \frac{H_0^2\Omega_m^0(1+z)^3}{H^2(z)}
\label{matden}
\end{eqnarray}
so that $\Omega_X(z)  = \Big{[}H_0^2\ \Omega_X^0\Big{/}H^2(z)\Big{]} \exp\left(\int_0^{z} 3(1+\omega_X(z))\frac{dz}{1+z}\right) = 1 - \Omega_m(z)$. The matter density parameter $\Omega_m(z)$ as given  in Eq.\ (\ref{matden}) can be written in terms of the parameters $a$, $b$ and $\Omega_m^0$ using the expression for $H(a,b;z)$ as
\begin{eqnarray}
\Omega_m(a,b,\Omega_m^0; z) &=& \Omega_m^0\frac{(1+z)^3(1+2az+(ab+a-b)z^2)^2}{[(1+z)(1+bz)]^4}
\label{omegamab}
\end{eqnarray}
With best-fit values of the parameters  $a$, $b$ and $\Omega_m^0$ obtained from analysis of observational data we use Eq.\ (\ref{omegamab}) to compute the $z$-variation of the density parameters both for matter ($\Omega_m$) and dark energy ($\Omega_X = 1 - \Omega_m$).

\section{Methodology of data analysis}
\label{sec:analysis}
The SNe Ia data remains the key observational ingredient in determining cosmological parameters related to dark energy. In this paper we have considered different compilations of SNe Ia observations including the recent UNION2 data  \citep{Amanullah}. The other SNe Ia data sets considered here are \citep{Riess07, WoodVasey, Davis, kowalski,Kessler}. These different groups tabulated the values of the distance modulus $\mu(z)$ for different values of the redshift $z$ from the SNe Ia observations. The distance modulus $\mu$ is related to the luminosity distance by 
\begin{eqnarray}
\mu(z) = 5\log\left[\frac{d_L(z)}{\rm 1~Mpc} \right] + 25 = 5\log\Big{[}D_L(z)\Big{]} + \mu_0\,\,,
\end{eqnarray}
where $D_L(z) = H_0 d_L(z)/c$ is the Hubble free luminosity distance and $\mu_0=42.38-5\log_{10}h$, with $h$ being a dimensionless parameter defining the value of the Hubble parameter at the present epoch as $H_0 = 100 h$ km s$^{-1}$ Mpc$^{-1}$. Using the parametric form of the $d_L$ (Eq.\ (\ref{dlzparam})), the distance modulus can be expressed in terms of the parameters $a$ and $b$ as 
\begin{eqnarray}
\mu_{\rm th}(a,b;z) =  5\log\left[ \frac{z(1+az)}{1+bz}\right] + \mu_0\,\,.
\end{eqnarray}
The observed values of the distance modulus $\mu_{\rm obs}(z_i)$ corresponding to  measured redshifts $z_i$ are given in terms of the absolute magnitude $M$ and the apparent magnitudes $m_{\rm obs}(z_i)$ by 
\begin{eqnarray}
\mu_{obs}(z_i)=m_{obs}(z_i)-M\,.
\end{eqnarray} 
To obtain the best-fit values of the parameters $a$ and $b$ from SNe Ia observations we perform a $\chi^2$ analysis which involves minimization of suitably chosen $\chi^2$ function with respect to the parameters $a$ and $b$. For our analysis of SNe Ia data we use the $\chi^2$ function considered  in \citep{Xu}. We refer the reader to \cite{Xu} for a comprehensive discussion on the choice of $\chi^2$ function and its minimization. The $\chi^2$ function (for the analysis of SNe Ia data) is first defined in terms of parameters $a$, $b$ and $M^\prime\equiv \mu_0 +M$ (called the nuisance parameter) as
\begin{eqnarray}
\chi^2_{\rm SN}(a,b,M^{\prime})
&=&\sum_{i=1}^N\frac{(\mu_{obs}(z_i)-
\mu_{th}(a,b,z))^2}{\sigma_i^2}  \nonumber\\
&=&\sum_{i=1}^N\frac{(5\log_{10}(D_L(a,b,z_i))-m_{obs}(z_i)+M^{\prime})^2}{\sigma_i^2}
\end{eqnarray}
where $\sigma_i$ is the uncertainty in observed distance modulus and $N$ is the total number of data points. Its marginalization over the nuisance parameter as $\bar{\chi}_{\rm SN}^2(a,b) = -2\ln\int_{-\infty}^{\infty} \exp\left[-\frac{1}{2} \chi^2(a,b,M^{\prime})\right]dM^{\prime}$ leads to $\bar{\chi}_{\rm SN}^2(a,b)=P - (Q^2/R)+\ln (R/2\pi)$ with
\begin{eqnarray*}
P &=&\sum_{i=1}^N\frac{(5\log_{10}(D_L(a,b,z_i))-m_{obs}(z_i))^2
}{\sigma_i^2} \\
Q &=&\sum_{i=1}^N\frac{(5\log_{10}(D_L(a,b,z_i))-m_{obs}(z_i))
}{\sigma_i^2} \quad {\rm and~~} \\
R &=&\sum_{i=1}^N\frac{1}{\sigma_i^2}\,\,.
\end{eqnarray*}
The function $\chi^2(a,b,M^\prime)$ has a minimum at $M^\prime=Q/R$ which gives the corresponding value of $h$ as $10^{(M-M^\prime+42.38)/5}$. Dropping the constant term $\ln (R/2\pi)$  from $\bar{\chi}^2_{\rm SN}$, the function
\begin{eqnarray}
\chi^2_{SN} (a,b) &=& P - \frac{Q^2}{R}
\label{chifinal}
\end{eqnarray}
can be used in for the likelihood analysis.

The observation of baryon acoustic oscillations provides another evidence for the existence of dark energy. In the early universe, free electrons and protons were coupled with highly energetic photons of the relativistic plasma through scattering. The high pressure in the plasma drives the primordial cosmological fluctuations to propagate outward at a relativistic speed. After the universe had cooled down sufficiently, at a certain point, electrons and protons combine to form neutral hydrogen atoms and thereby decoupling photons from baryons. This causes an abrupt fall of the speed of propagation of the acoustic wave. The baryon acoustic oscillation leaves their signature on the large scale structure of the universe. The Slogan Digital Sky Survey (SDSS) measures the correlation function of the large sample of luminous red galaxies. The acoustic peak detected by them provides a standard ruler with which the absolute distance corresponding to a typical redshift $z=0.35$ can be determined. The standard ruler is given by the dimensionless parameter $A$ in Eq.\ (\ref{constructedA}) as $A = 0.469 \pm 0.017$. The dimensionless quantity $A$ is as given in Eq.\ (\ref{constructedA}) is constructed from the following set of equations
{\small
\begin{eqnarray}
A(z) = D_V(z)\frac{\sqrt{\Omega_m^0 H_0^2}}{z} ,
D_V(z)={\left[\frac{D_A^2(z) z}{H(z)}\right]}^{1/3} ,
D_A(z)=\int_0^z \frac{dz^\prime}{H(z^\prime)}\label{setA}
\end{eqnarray}
}
where $D_A(z)$ is the comoving angular diameter distance and $D_V(z)$ is the dilation scale. Considering $a$, $b$ and $\Omega_m^0$ as parameters to be determined from observations, the $\chi^2$ function for analysis of BAO data is given as
\begin{eqnarray}
\chi^2_{\rm BAO}(\Omega_m^0,a,b) &=& \frac{[A(\Omega_m^0,a,b) - A_{\rm obs}]^2}{(\Delta A)^2}
\label{chibao}
\end{eqnarray}
with $A_{\rm obs} = 0.469$, $\Delta A = 0.017$ and $A(\Omega_m^0,a,b)$ being given by Eq.\ (\ref{Aabomega}).

Determination of Hubble parameter from observational measurements is another probe to the accelerated expansion of the universe attributed to the dark energy. Compilation of the observational data based on measurement of differential ages of the galaxies by Gemini Deep Deep Survey GDDS \citep{Abraham}, SPICES  and VDSS surveys provide the values of the Hubble parameter at 15  different redshift values \citep{x39,x40,x41,mth20}. The $\chi^2$ function for the analysis of this observational Hubble data can be defined as
\begin{eqnarray}
\chi^2_{\rm OHD}(a,b) &=& \sum_{i=1}^{15} \left [ 
\frac {H(a,b;z_i) - H_{\rm obs}(z_i)} {\Sigma_i} \right ]^2 \,\,,
\end{eqnarray}
where $H_{\rm obs}$ is the observed Hubble parameter value at $z_i$ with uncertainty $\Sigma_i$.

Varying the parameters $a$, $b$ and $\Omega_m^0$ freely we minimize the $\chi^2$ function which is defined as 
\begin{eqnarray}
\chi^2(a,b,\Omega_m^0) &=& \chi^2_{\rm SN}(a,b) + \chi^2_{\rm BAO}(a,b,\Omega_m^0) +  \chi^2_{\rm OHD}(a,b)\,\,.
\end{eqnarray}
The values of the parameters $a$, $b$ and $\Omega_m^0$ at which minimum of $\chi^2$ is obtained are the best-fit values of these parameters for the combined analysis of the observational data from SNe Ia, BAO and OHD. With these values of the parameters we find the variation of the dark energy equation of state $w_X(z)$ and the dark energy density parameter
($\Omega_X(z) = 1 - \Omega_m(z)$) using Eqs.\ (\ref{wxab}) and (\ref{omegamab}) respectively. We also find the 1$\sigma$ ranges  of the parameters $a$, $b$ and $\Omega_m^0$ from the analysis of the observational data discussed above. In this case of three parameter fit, the  1$\sigma$ (68\% confidence level) allowed ranges of the parameters correspond to $\chi^2 \leq \chi^2_{\rm min} + \Delta\chi^2$, where $\Delta\chi^2(=3.53)$ denotes the 1$\sigma$ spread in $\chi^2$ corresponding to three parameters. For these allowed domains of the parameters, we also obtain the corresponding 1$\sigma$ ranges of the quantities $w_X(z)$, $\Omega_X(z)$ and $\Omega_m(z)$ from Eqs.\ (\ref{wxab}) and (\ref{omegamab}). The computation of $\Omega_m(z)$ from Eq.\  (\ref{omegamab}) with the parameters $a$, $b$ and $\Omega_m^0$ as inputs from their 1$\sigma$ ranges  obtained in a way described above does not directly ensure that the condition $\Omega_m(z) \leq 1$, which follows from the definition of $\Omega_m(z)$, is always respected. (However, we have seen that for the best-fit values of parameters $a$, $b$ and $\Omega_m^0$ as obtained from the analysis, $\Omega_m(z)$ lies below 1 for the range of $z$ probed by SNe Ia observations.) To circumvent this, $\Omega_m(z)$ at some particular value of $z$ corresponding to an earlier epoch (beyond the range of measured redshifts $z \la 1.76$ of SNe Ia events) is not allowed to exceed some chosen benchmark value (say, $\alpha$) below 1. To take into account this constraint we find the domain of the ($a , b , \Omega_m$) parameter space for which the two conditions, \textit{viz.} $\chi^2 \leq \chi^2_{\rm min} + \Delta\chi^2$ and $\Omega_m(a,b,\Omega_m^0;z) \leq \alpha$ are simultaneously satisfied. The 1$\sigma$ range of the parameters thus obtained are, therefore, dependent on the initial condition of matter density at some earlier epoch which we choose here as $z=2$. We study the impact of imposing the constraint ($\Omega_m(z=2) \leq \alpha$)  on the parameter space by finding allowed domains of the parameter space for different choices of values of $\alpha$. We also study this effect on the allowed region of the equation of state ($w_X(z)$) of dark energy.

\section{Results and discussions}
\label{sec:results}
In this section we present and discuss the results of combined analysis of the SNe Ia, OHD and BAO data. As mentioned earlier we have considered five different sets of data for the analysis of observations of Supernova Ia along with the OHD and BAO data. The SNe Ia data sets considered here are  HST+SNLS+ESSENCE \citep{Riess07, WoodVasey, Davis}, SALT2 data and MLCS data \citep{Kessler}, UNION data \citep{kowalski} and UNION2 data  \citep{Amanullah}. In the analysis, we have taken each one of these five sets of SNe Ia data at a time with OHD and BAO data to compute $\chi^2 = \chi^2_{\rm SN} + \chi^2_{\rm OHD} + \chi^2_{\rm BAO}$ for different sets of values of the parameters $a$, $b$ and $\Omega_m^0$. 

{\small
\begin{table}[h]
\begin{center}
\begin{tabular}{ccc}
\hline
SNe Ia data sets &    best-fit values of &   Minimum \\
+ BAO + OHD      &  ($a$, $b$, $\Omega_m^0$) & value of $\chi^2$\\ 
\hline
HST+SNLS+ESSENCE  &  & \\
+ BAO + OHD     & (1.437, 0.550, 0.268) & 199.267\\
(Data Set: I)       &&\\
No. of data points = 192+1+15 &&\\
\hline
SALT2  & & \\
+ BAO + OHD      & (1.401, 0.542, 0.272) & 560.083\\
(Data Set: II)      &&\\
No. of data points = 288+1+15  &&\\
\hline
MCLS &  & \\
+ BAO + OHD       & (1.401, 0.653, 0.296) & 783.078\\
(Data Set: III)   &&\\
No. of data points = 288+1+15  &&\\
\hline
UNION  & & \\
+ BAO + OHD     & (1.635, 0.699, 0.268 ) & 311.615\\
(Data Set: IV)    &&\\
No. of data points = 307+1+15 &&\\
\hline
UNION2 & & \\
+ BAO + OHD      & (1.289, 0.458, 0.272) & 544.074\\
(Data Set: V)    &&\\
No. of data points = 557+1+15 &&\\
\hline
\end{tabular}
\caption{\label{tab1} Best-fit values of parameters and minimum values of $\chi^2$ for each of the 5 data sets considered.}
\end{center}
\end{table}
}
\begin{figure}[h]
\includegraphics[height=3cm,width=8cm,angle=0]{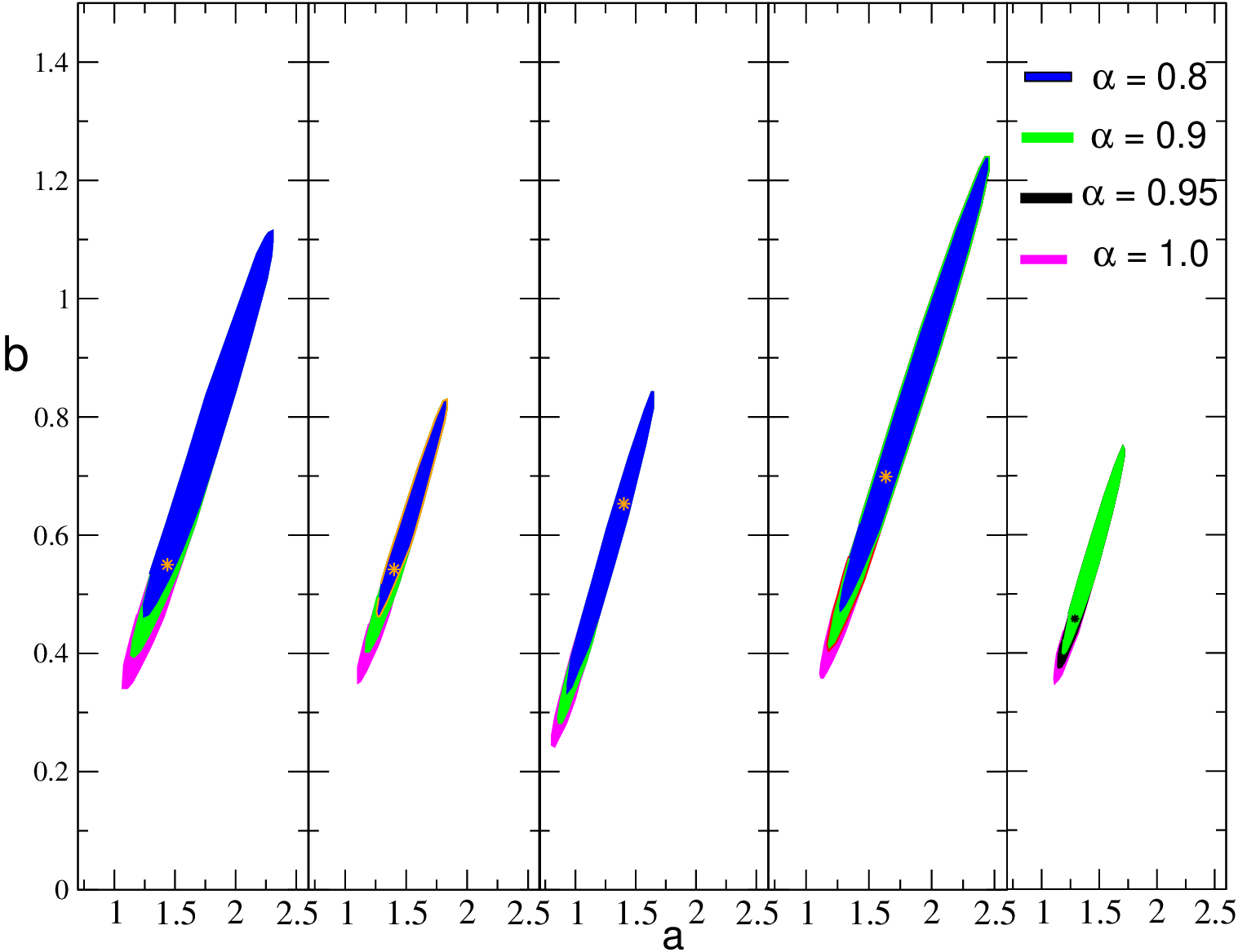}
\caption{\label{fig:1}1$\sigma$ contour in the $a$ - $b$ parameter space (with marginalization over $\Omega_m^{(0)}$) obtained from analysis of  data sets I-V (see text) from left to right. In each of the 4 figures from left, contours are shown for three different values of $\alpha=0.8, 0.9\ {\rm and}\ 1$. The figure in the extreme right (for data set V) contours are shown for $\alpha=0.9, 0.95\ {\rm and}\ 1$ (described in text).}
\vskip .5cm
\includegraphics[height=3cm,width=8cm,angle=0]{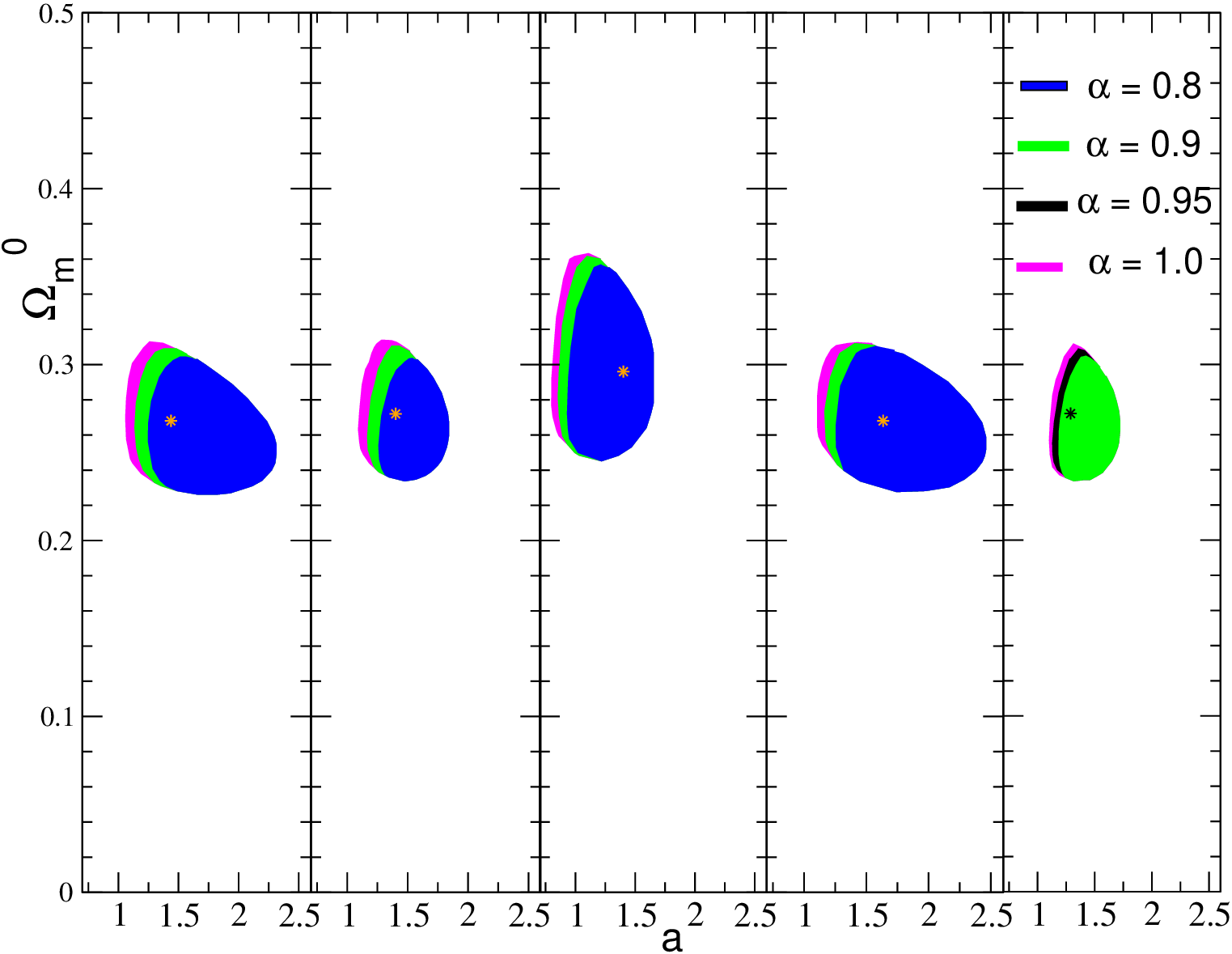}
\caption{\label{fig:2}1$\sigma$ contour in the $a$ - $\Omega_m^0$ parameter space (with marginalization over $b$) obtained from analysis of  data sets I-V (see text) from left to right for same set of values of $\alpha$ as in Figure\ \ref{fig:1}.}
\vskip .5cm
\includegraphics[height=3cm,width=8cm,angle=0]{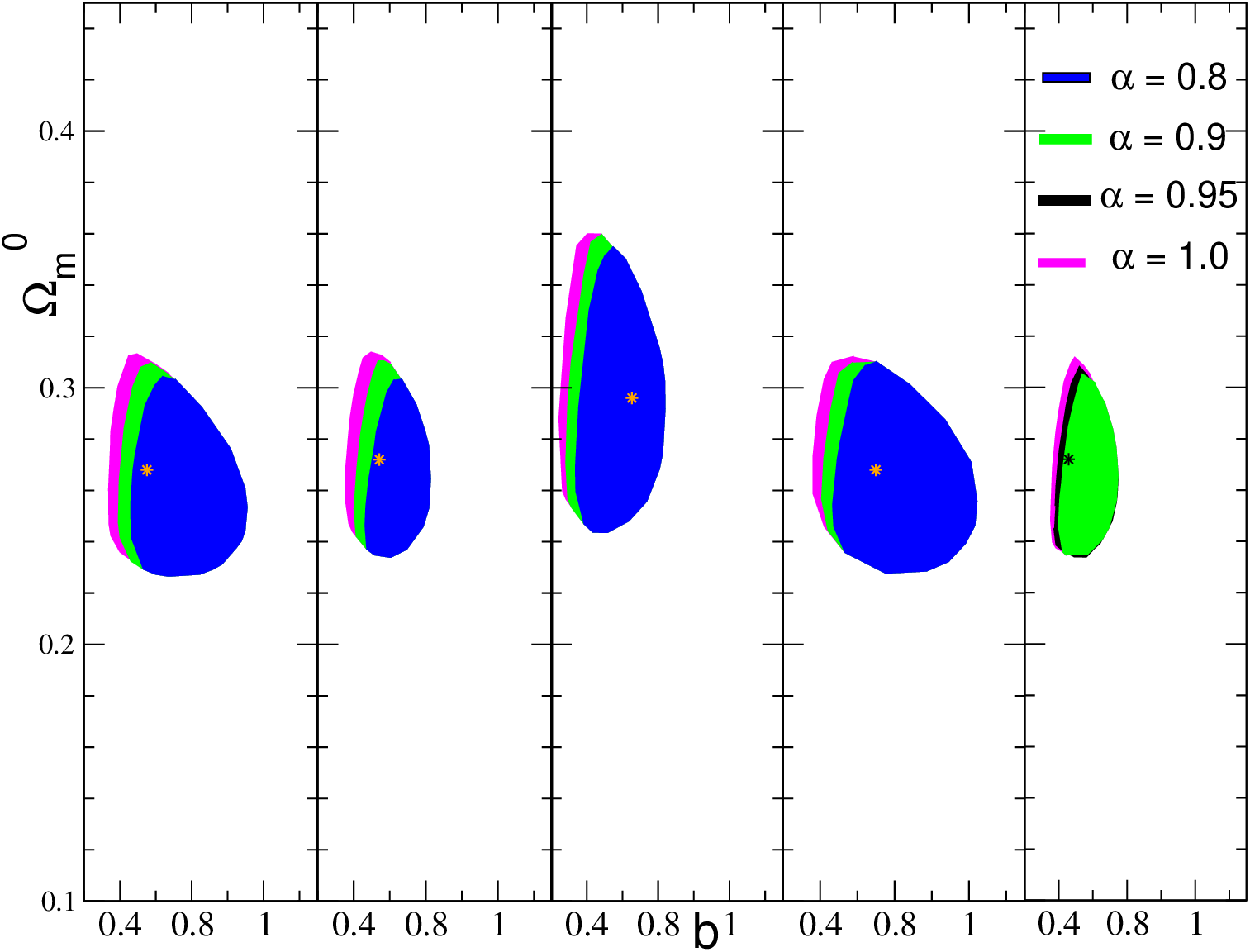}
\caption{\label{fig:3}1$\sigma$ contour in the $b$ - $\Omega_m^0$ parameter space (with marginalization over $a$) obtained from analysis of  data sets I-V (see text) from left to right for same set of values of $\alpha$ as in Figure\ \ref{fig:1}.}
\end{figure}
In Table\ \ref{tab1} we present the best-fit values of the parameters $a$, $b$ and $\Omega_m^0$ obtained from analysis of different data sets. The minimum value of $\chi^2$ along with total number of data points for each set of data are also shown. We refer to the 5 different sets of data considered here by `set I', `set II', `set III', `set IV' and `set V' as accordingly listed in column 1 of Table\ \ref{tab1}. We then find the $1\sigma$ ranges of the parameters $a$, $b$ and $\Omega_m^0$ for each of the data sets (I-V) using the method described in Sec.\ \ref{sec:analysis}. To ensure the boundedness: $\Omega_m(z)<1$ , we obtain the allowed ranges of the parameters for different values of $\alpha$ (defined in Sec.\ \ref{sec:analysis}). The 1$\sigma$ allowed region of the parameters obtained from the combined analysis of SNe Ia, OHD and BAO data are presented in  the planes of any two parameters of set \{$a$,  $b$, $\Omega_m^0$\} by marginalizing over the third one. In Figures\ \ref{fig:1}, \ref{fig:2} and \ref{fig:3} we show the 1$\sigma$ contours in parameter planes $a-b$, $a-\Omega_m^0$ and $b-\Omega_m^0$ respectively for different choices of the values of $\alpha$ ($\Omega_m(z=2)\leq \alpha$). The five  panels (row-wise) from left to right in each of the Figures (Figure\ \ref{fig:1}, \ref{fig:2} and \ref{fig:3}) correspond to results of analysis of data sets I to V. For panels 1-4 (corresponding to data sets I-IV) in these three figures, the 1$\sigma$ contours are plotted for three different values of $\alpha$ \textit{viz.} 0.8, 0.9 and 1.0 \textit{i.e.} for $\Omega_m(z=2) \leq 0.8, 0.9\ {\rm and}\ 1$. In the last panel corresponding to the data set V (UNION2 data along with OHD and BAO), the same is plotted for $\alpha=0.9,  0.95\ {\rm and}\ 1$. The values of $\alpha$ below 0.9 are not chosen for data set V (UNION2+OHD+BAO) because  the value of $\Omega_m(z=2)$ exceeds 0.9 even when calculated at the best-fit values of parameters $a$, $b$ and $\Omega_m^0$ obtained from the analysis of data set V. The  UNION2 data (along with OHD and BAO) thus restricts  the matter density parameter value at an epoch $z=2$ to lie slightly below $0.95$ (at 1$\sigma$ level).
\begin{figure}[t]
\includegraphics[height=5cm,width=8.5cm,angle=0]{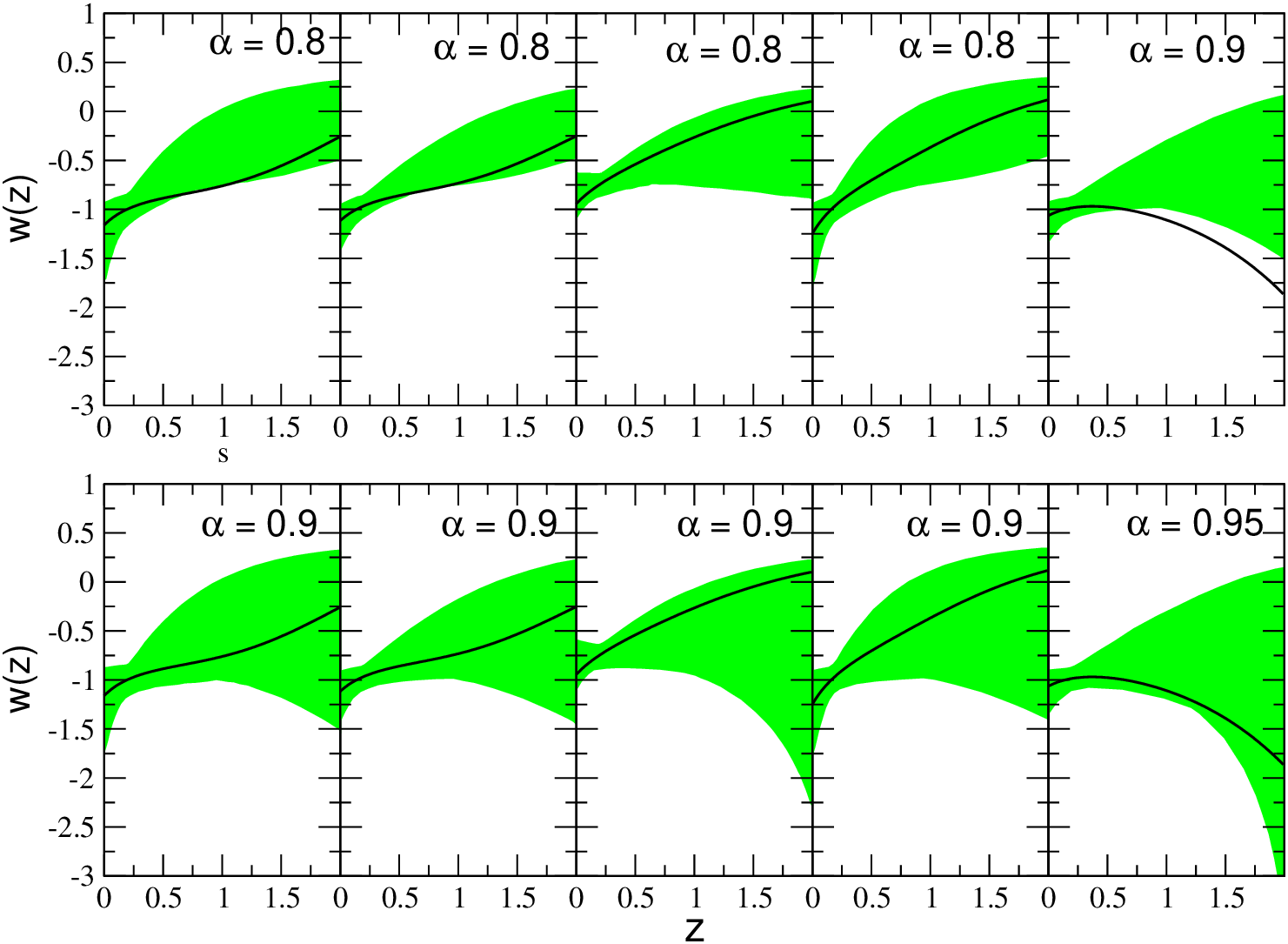}
\caption{\label{fig:4} Plots of $w_X(z)$ vs $z$ for different data sets : Columns from left to right correspond to data sets I-V respectively. Figures for data set I-IV are plotted for $\Omega_m (z=2)\leq 0.8$ (upper panel) and  $\Omega_m (z=2)\leq 0.9$ (lower panel). For data set V (UNION2), plots are shown for $\Omega_m (z=2) \leq 0.9$ and 0.95 at upper and lower panels respectively.}
\vskip .5cm
\includegraphics[height=5cm,width=8.5cm,angle=0]{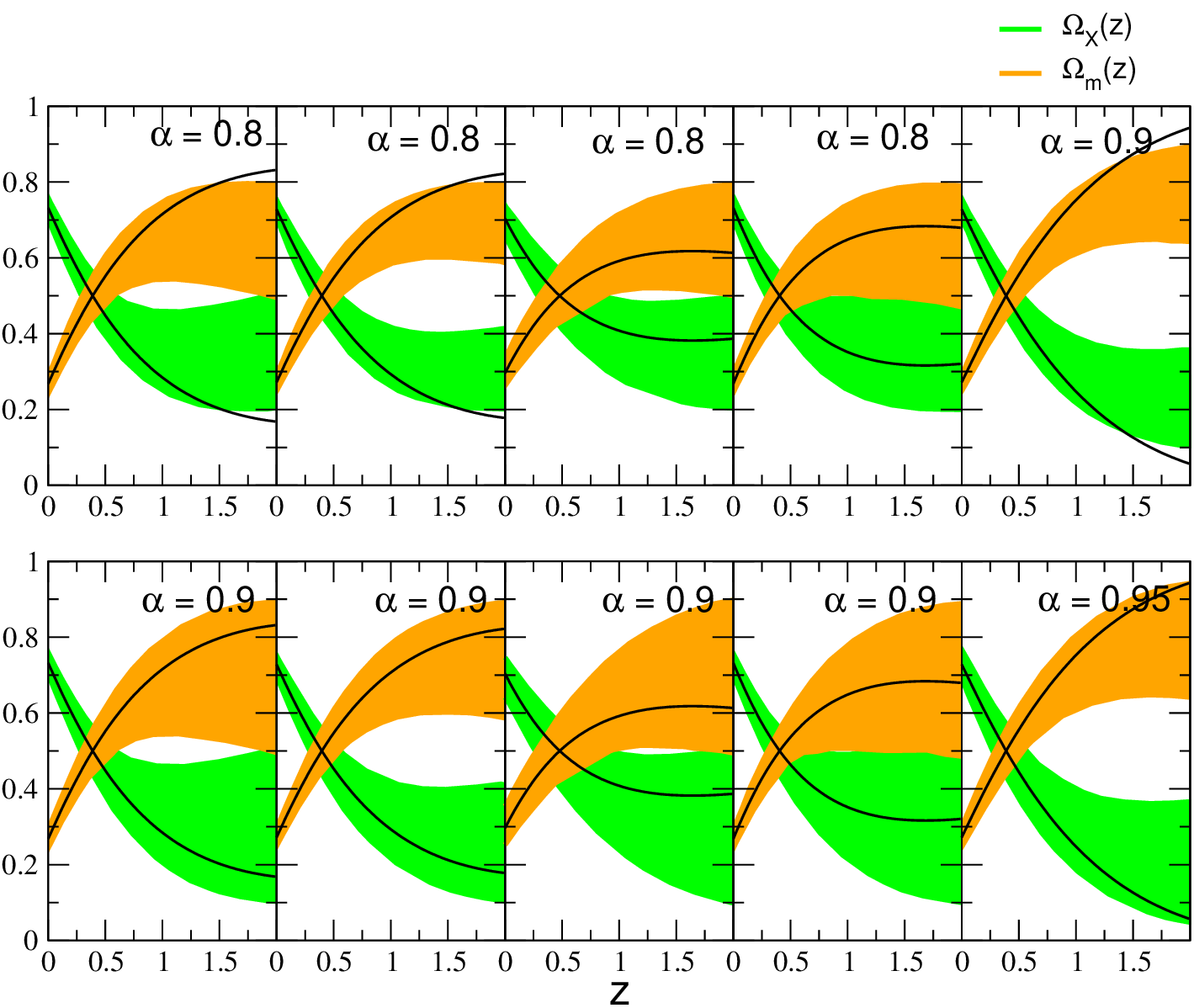}
\caption{\label{fig:5} Plots of $\Omega_X(z)$ and $\Omega_m(z)$ vs $z$ for different data sets : Columns from left to right correspond to data sets I-V respectively. Figures for data set I-IV are plotted for $\Omega_m (z=2)\leq 0.8$ (upper panel) and  $\Omega_m (z=2)\leq 0.9$ (lower panel). For data set V (UNION2), plots are shown for $\Omega_m (z=2) \leq 0.9$ and 0.95 at upper and lower panels respectively.}
\end{figure}

With the best-fit values of the parameters $a$, $b$ and $\Omega_m^0$ as obtained above (listed in Table\ \ref{tab1} for different data sets) we compute the equation of state $w(z)$ of dark energy as a function of redshift $z$ using Eq.\ (\ref{wxab}). The plots for $w(z)$ vs $z$ are shown by solid curves in Figure\ \ref{fig:4} for all the 5 data sets. Plots from 1-5 (row-wise) correspond to data sets I-V. Using the 1$\sigma$ range for the parameters $a$,$b$ and $\Omega_m^0$ as obtained above for different values of $\alpha\ (\Omega_m(z) \leq \alpha)$ we obtain the corresponding spread in $w(z)$. The shaded regions  in Figure\ \ref{fig:4} show these 1$\sigma$ bands of $w(z)$. This has been shown for different choices of the value of $\alpha$. The first 4 plots from left to right in the upper \textbf{(}lower\textbf{)} panel are for data sets I-IV respectively with $\Omega_m(z=2) \leq 0.8$ \textbf{(}$\Omega_m(z=2) \leq 0.9$\textbf{)}. The two plots in the extreme right column correspond to the data set V with the two constraints: $\Omega_m(z=2) \leq 0.9$ (upper panel) and $\Omega_m(z=2) \leq 0.95$ (lower panel). The same best-fit curves for $w_X(z)$ vs $z$ are plotted both in the upper and lower panels of a given column. From Figure\ \ref{fig:4} we observe that, in some cases, the $w_X(z)$ vs $z$ plots (solid curves) corresponding to the best-fit values of the parameters $a$, $b$ and $\Omega_m^0$ lie well within the respective 1$\sigma$ regions. They barely remain within  such regions in some other cases. In one other occasion, the best-fit plot is outside the 1$\sigma$ region for most of the range of $z$ considered. For example, for data sets III and IV the best-fit curve remain within the corresponding 1$\sigma$ range for both the choices:  $\Omega_m(z=2) <0.8$ (upper panels, columns 3 \& 4) and  $\Omega_m(z=2) <0.9$ (lower panel, columns 3 \& 4). For data sets I and II as shown respectively in first and second columns of Figure\ \ref{fig:4}), however, one can see that although the best-fit lines are within the 1$\sigma$ regions obtained for $\Omega_m(z=2) \leq 0.9$ but certain segments of the best-fit lines tend to come out of the respective 1$\sigma$ domains obtained for the case $\Omega_m(z=2) \leq 0.8$.  In the last column of Figure\ \ref{fig:4}, corresponding to data set V, we show the 1$\sigma$ spread in the variation of $w_X(z)$ with $z$ for the two choices: $\Omega_m(z)<0.9$ (upper panel) and  $\Omega_m(z)<0.95$ (lower panel). As evident from the plots, a large segment of the best-fit $w_X(z)$-$z$ curve in this case is not contained within the 1$\sigma$ region obtained for the case $\Omega_m(z=2) \leq 0.9$ where as fully contained within the 1$\sigma$ region  obtained for $\Omega_m(z=2) \leq 0.95$.

The above results can be interpreted and summarized like this: The data sets I ((HST+SNLS+ESSENCE)+BAO+OHD) and II (SALT2+BAO+OHD) support the fact that matter density parameter $\Omega_m$ at an early stage of the universe at $z=2$ was $\ga$ 0.8 whereas the data sets III (MLCS+BAO+OHD) and IV (UNION+BAO+OHD) can accommodate values of matter density parameter at the epoch $z=2$ even a bit lower than 0.8. According to analysis of data set V (UNION2+BAO+OHD), the matter density parameter at $z=2$ is only allowed to have values greater than 0.9.

Another important observation in the context of Figure\ \ref{fig:4} is the following. The variation of $w_X(z)$ over the range $0 \la z \la 1.76$ (relevant for SNe Ia) with its 1$\sigma$ spread obtained from data sets I, II and IV with $\Omega_m(z=2) \leq 0.8$ (columns 1, 2 and 4 in the upper panel of Figure\ \ref{fig:4} show non-overlap between the 1$\sigma$ bands of $w_X(z)$ at two well separated $z$ values in the range $0 \la z \la 1.76$. This phenomenon  points to observational evidence (at 1$\sigma$ level) for varying equation of state of dark energy as opposed to cosmological constant solution. On the contrary, such signatures of varying dark energy are not obtained from the analysis of the sets of data I, II, IV and V with $\Omega_m(z=2) \leq 0.9$. This is evident from the plots in the lower panels of columns 1, 2 and 4 and in the upper panel of column 5 of Figure \ref{fig:4}. The other data set III also does not provide any signature of varying dark energy (at 1$\sigma$ level) even for $\Omega_m(z=2) \leq 0.8$. Therefore, knowledge of the matter density of the universe at some earlier epoch is instrumental in  providing observational evidences in favor of varying dark energy or cosmological constant solutions.   

Using Eq.\ (\ref{omegamab}) we compute the matter density parameter $\Omega_m(z)$ as a function of $z$ for the best-fit values of the parameters $a$, $b$ and $\Omega_m^0$. The 1$\sigma$ range of the parameter $\Omega_m(z)$ for different choices of the value of $\alpha$ are also computed. The results are shown in Figure\ \ref{fig:5} by solid lines and shaded regions respectively. The plots for dark energy density parameter $\Omega_X(z) = 1 - \Omega_m(z)$ are also shown. Columns 1 to 5 respectively correspond to data sets I-V. For the upper and lower panels of columns 1-4, imposed constraints on $\Omega_m(z=2)$ are $\Omega_m(z=2) \leq 0.8$ and $\Omega_m(z=2) \leq 0.9$ respectively. For column 5 plots are shown for $\Omega_m(z=2) \leq 0.9$ (upper panel) and $\Omega_m(z=2) < 0.95$ respectively.  The dark energy starts dominating at the value of $z$ where plots for $\Omega_m(z)$ and $\Omega_X(z)$ intersect. The estimations of such an epoch for each of the data sets considered may be obtained from the point of intersections of the corresponding best-fit plots of $\Omega_m(z)$ and $\Omega_X(z)$. For example,  this intersection point are obtained at $z \simeq 0.4$  for data sets I,II, IV and V whereas for data sets III the 2 plots intersect at $z \simeq 0.48$. As in Figure\ \ref{fig:4}, in this figure also, the best-fit plots for data set I and II for the choice $\Omega_m(z=2) = 0.8$  do not remain contained within the corresponding $1 \sigma$ regions. The same is true for the data set V (upper panel of the last column of Fig. 5; $\Omega_m(z=2) = 0.9$).  From Figure\ \ref{fig:5} we also obtain the $1 \sigma$ uncertainty for $\Omega_m^0$ - the matter density parameter at present epoch. For data sets I, II, IV and V the $1 \sigma$ spread of  $\Omega_m^0$ is $\sim 0.23 - 0.31$  and for data set III this spread is $\sim 0.25 - 0.35$. These ranges are calculated with the choice $\Omega_m(z=2) = 0.8$ for  data sets I,II,III and IV while for data set V the estimation is for the choice $\Omega_m(z=2) = 0.95$. We like to comment that these ranges will be affected negligibly for other choices of $\Omega_m(z=2)$ considered here as is obvious from Figures.\ \ref{fig:2}  and \ref{fig:3}. The earlier discussed facts that the data sets I and II allow  matter density parameter values at an  early stage of the universe at z = 2 $\ga$ 0.8 whereas data sets III and IV can accommodate even lower values of matter density parameter at the epoch $z=2$ and  data set V only allows  values a bit greater than 0.9 are also reflected in the plots of Figure\ \ref{fig:5} through the containment or non-containment of the best-fit plots within the corresponding 1$\sigma$ limits.

\section{Conclusion}
\label{sec:conclusion}
In this paper we have performed a combined analysis of the SNe Ia, OHD and BAO data assuming a closed form parametrization of the luminosity distance in terms of redshift. Five different sets of SNe Ia data designated as (HST+SNLS+ESSENCE), SALT2, MLCS, UNION and UNION2 are independently analysed combining the individual data sets with BAO and OHD. Among the 5 SNe Ia data sets, the UNION2 data is the most recent. From the analysis we find the best-fit values of the parameters $a$ and $b$ appearing in the luminosity distance - redshift parametrization along with the matter density parameter at the present epoch ($\Omega_m^0$). For this we use a 3-parameter $\chi^2$ fit to each of the SNe Ia data sets along with OHD and BAO. We also find the 1$\sigma$ ranges of the parameters $a$, $b$ and $\Omega_m^0$  imposing a constraint on the matter density parameter ($\Omega_m$) that at earlier epoch (chosen here as $z=2$) its value does not exceed a certain value $\alpha$. Results are presented for some benchmark values of $\alpha$ as 0.8, 0.9, 0.95 etc. We also compute $z$-variations of the equation of state of dark energy ($w_X(z)$) and matter and dark energy density parameters ($\Omega_m(z)$ and $\Omega_X(z)$) corresponding to the best-fit values of the parameters $a$, $b$ and $\Omega_m^0$ and their 1$\sigma$ ranges.  The results of the analysis show that  knowledge of the matter density of the universe at some earlier epoch is instrumental in providing observational evidences in favor of varying dark energy or cosmological constant solutions. Also the SNe Ia data sets (HST+SNLS+ESSENCE) and SALT2 (along with BAO and OHD) restrict the matter density parameter value at the earlier epoch ($z=2$) not to go below 0.8. The data sets MLCS and UNION1, on the other hand restrict the same, above a value which is slightly greater than 0.8. The recent UNION2 data, however, constrains the value to lie always above 0.9. We obtain the matter density parameter at present epoch, $\Omega_m^0 \sim 0.27$ for four sets of data (I, II, III and V) while for data set IV, $\Omega_m^0 \sim 0.3$. The 1$\sigma$ spread for  $\Omega_m^0$  lies within the range $\sim 0.23 - 0.31$ for data sets I, II, IV and V whereas for the data set III this range is $\sim 0.25 - 0.35$.  The nature of variations of the dark energy equation of state are similar for data sets I-IV and is different for the data set V. The analysis of the data sets I, II, IV and V show that the dark energy starts dominating the matter from the epoch $z \sim 0.4$ and the same from the analysis of data set V is found to be $z \sim 0.48$.

\section*{Acknowledgements}
The authors thank Anjan Ananda Sen for some useful discussions.

\end{document}